\documentclass[preprint,prb,showpacs,endfloats]{revtex4}

\begin{document}
\title{Work functions, ionization potentials, and in-between:
\linebreak Scaling relations based on the image charge model}
\author{Kin Wong}
\author{Sascha Vongehr}
\author{Vitaly V. Kresin}
\affiliation{ Department of Physics and Astronomy, University of
Southern California, Los Angeles, CA 90089-0484, USA }
\date{\today}

\begin{abstract}
We revisit a model in which the ionization energy of a metal particle
is associated with the work done by the image charge force in moving
the electron from infinity to a small cut-off distance just outside
the surface.  We show that this model can be compactly, and
productively, employed to study the size dependence of electron
removal energies over the range encompassing bulk surfaces, finite
clusters, and individual atoms.  It accounts in a straightforward
manner for the empirically known correlation between the atomic
ionization potential (IP) and the metal work function (WF),
IP/WF$\sim$2.  We formulate simple expressions for the model
parameters, requiring only a single property (the atomic
polarizability or the nearest neighbor distance) as input. Without
any additional adjustable parameters, the model yields both the IP
and the WF within $\sim$10\% for all metallic elements, simulates the
concentration dependence of the WF of regular binary bulk alloys, and
matches the size evolution of the IP of finite metal clusters for a
large fraction of the experimental data. The parametrization takes
advantage of a remarkably constant numerical correlation between the
nearest-neighbor distance in a crystal, the cube root of the atomic
polarizability, and the image force cutoff length. The paper also
includes an analytical derivation of the relation of the outer radius
of a cluster of close-packed spheres to its geometric structure.
\end{abstract}

\pacs{79.60.Jv, 78.67.-n, 61.46.+w} \maketitle

\section{INTRODUCTION}
\label{sec:intro}

While the good agreement between theoretical and experimental atomic
ionization potentials (IP) is a major triumph for quantum mechanics,
it is prohibitively more difficult to rigourously solve the
polyatomic quantum problem, not to mention extrapolation to an
infinite bulk metal. The IP for an atom is a well understood
quantity. The same cannot, however, be said about the work function
(WF) for a metal or even for a finite cluster. On the other hand, it
is an experimentally realistic task to produce clusters ranging in
size from two atoms up to tens of thousands of atoms and to measure
the size dependence of the electron removal energy. The clusters can
be made big enough that they can be considered close to bulk metals,
hence the evolution from the atomic IP to the metal WF can be mapped
out. Experiments over such a wide range have been performed, e.g.,
for sodium\cite{gohlich:91}. For each element in the periodic table
one would therefore expect that there exists a function which can
predict the electron removal energy for a particle of arbitrary size.

The exact derivation of such a scaling law is a daunting task.
However, the available experimental data on the IP, the WF, and on
clusters reveal some characteristic features. For example, it has
been noted a long time ago that the IP and the WF of metallic
elements are approximately correlated to each other via the
factor\cite{gallo:74,rose:83}
\begin{equation}
\label{IPWF} I/W \approx 2.
\end{equation}
To give another example, for many metal clusters of intermediate
sizes the electron removal energy has been found to scale as
\begin{equation}
\label{clusterIP} \phi \approx W + \gamma{e^2 \over R},
\end{equation}
where $R$ is the radius of the cluster and $\gamma$ is a constant
factor \cite{seidl:91,deheer:93}. Henceforth, $I$ will denote the
atomic first ionization potential, $W$ will denote the
polycrystalline bulk surface work function, and for a finite metal
particle the term "electron removal energy" will be employed
(denoted by $\phi$).

A significant number of first-principles calculations have been
performed for work functions of metallic
systems\cite{lang:83,jakobi:94,kiejna:96}. In addition, simple
models based on semi-empirical approaches combined with classical
electrostatics have been rather successful at reproducing WF
trends and values \cite{durak:01}. This suggests that some
features of the desired scaling law may be found by employing such
a model. In this paper we demonstrate that by combining the
image-charge potential function for a finite particle with just a
single material-dependent scale parameter (the atomic
polarizability $\alpha$ or the crystalline nearest-neighbor
distance $r_{nn}$) one can obtain an interpolation formula
covering the full size range from the atom through the cluster to
the bulk. This formula estimates both the IP and the WF within
$\sim$ 10\% for all metallic elements in the periodic table,
yields values in reasonable agreement with experiment for many
intermediate sized clusters, and provides a natural justification
for the aforementioned IP/WF$\approx$ 2 ratio.

The plan of the paper is as follows. In Sec. \ref{sec:image} we
consider the image-charge expression which describes the removal of
an electron from an isolated sphere. By focusing on the limits of a
sphere of atomic radius and one of infinite radius, we show that the
relation (\ref{IPWF}) follows as naturally from the image-charge
consideration as does Eq. (\ref{clusterIP}). In Sec. \ref{sec:result}
we demonstrate some striking parallels between the variation of
atomic sizes and image-force cutoff distances across the periodic
table and use these observations to formulate compact expressions for
estimating both WF and IP. An extension to the case of binary alloys
is discussed in Sec. \ref{sec:alloys}.  Finally, we show that by
using an appropriately interpolated expression for the cluster radius
a good description of the size evolution of cluster ionization
potentials can be obtained (Sec. \ref{sec:radius}). Some rigorous
formulas on close-packed cluster radii are derived in the Appendix.

\section{THE IMAGE POTENTIAL MODEL}
\label{sec:image}

It is stated in the literature that the earliest attempt to explain the
work function of a metallic surface using classical electrostatics
was due to Debye\cite{debye:1910}. He proposed that it is equal to
the energy required to pull an electron out to infinity against its
image charge. Since the image force diverges at the surface, Schottky
\cite{schottky:1923} suggested that one may be able to define a
microscopic cutoff distance $d$ at which the image force starts to
act. Despite the simplicity of this model\cite{density}, there have
been numerous attempts to estimate the parameter $d$ in order to fit
the experimental values of the WF
\cite{gallo:74,rose:83,brodie:95,halas:98,wojcie:99,durak:01}.
Although the particular choices of $d$ were supported only by
plausibility arguments, they were frequently able to offer rather
nice agreement with the experimental data.

In a similar spirit, let us now consider the image-force expression
for the energy required to remove an electron from an isolated finite
metal particle, modelled as a conducting sphere of radius $R$.  The
particle is assumed initially neutral, i.e., after the removal of the
electron it acquires an unit positive charge. A calculation of the
work required to move the electron from a distance $d$ outside the
metal surface to infinity against its image charge is a
straightforward exercise \cite{jackson}:
\begin{equation}
\label{IP(R)} \phi(R) = {e^2 \over 4d} \left( {1 + 4(d/R) + 6(d/R)^2
+ 2(d/R)^3 \over (1 + d/2R)(1 + d/R)^2} \right).
\end{equation}
The cutoff parameter $d$ is assumed to be a material-dependent
constant. The first factor on the right-hand side represents the bulk
($R \to \infty $) work function: $W={e^2 /(4d)}$. This formula has
been applied to yttrium and lanthanide clusters in Ref.
[\onlinecite{durak:chem:01}]. However, the authors did not extend it
to the bulk or the atomic limit; they used the equation as an
extrapolation formula for small clusters with the WF as a given
boundary condition.

It is convenient to rewrite the above equation in the following form:
\begin{equation}
\label{phi/WF} {\phi (R) \over W} = \eta (d/R),
\end{equation}
where the dimensionless scaling function $\eta (d/R)$ is defined by
the second factor in Eq. (\ref{IP(R)}) and plotted in Fig.
\ref{fig:scalingfunction}.
If Eq. (\ref{phi/WF}) were applied all the way down to the atomic
limit, it would give an estimate of the ratio IP/WF as the value of
$\eta (d/R_{at})$, where $R_{at}$ is a quantity characterizing the
atomic size. On the other hand, we know from numerous investigations
\cite{gallo:74,rose:83,brodie:95,halas:98,wojcie:99,durak:01} that
the cutoff parameter $d$ is, sensibly enough, also of the same order
of magnitude. In the atomic limit, therefore, the ratio $d/R_{at}$
should be on the order of unity. In other words, if one assumes that
$d \sim R_{at}$, then, independent of the exact expression for either
parameter, the scaling function predicts that
\begin{equation}
\label{IPWF2} I/W \sim \eta(1) \approx 2.
\end{equation}
This is a new explanation of the well known empirical result
mentioned in the Introduction, Eq. (\ref{IPWF}). In the next section
we suggest some specific parametrizations of the variables $d$ and $
R_{at}$ and show that these can give an even more accurate value of
the IP/WF ratio.

In the large $R$ limit, the scaling function can be expanded to first
order in $d/R$. The result is:
\begin{equation}
 \phi(R) = W + {3\over 8}{e^2\over R} + O\left(({d \over R})^2\right).
\label{finitesize}
\end{equation}
This is the well known finite size correction for the ionization
potential of metallic clusters
\cite{smith:65,wood:81,seidl:91,deheer:93,brechignac:94}. This
scaling law has been experimentally verified for many metal clusters
\cite{seidl:91,dugourd:92,deheer:93,brechignac:94}. Although there is
still some controversy whether the 3/8 factor is sufficiently
rigorous\cite{makov:88,deheer:90,harbola:92,seidl:94,yannouleas:96},
Eq. (\ref{finitesize}) does fit the experimental data relatively
well.

It appears, therefore, that Eq. (\ref{IP(R)}) offers a consistent
estimate for the scaling of electron removal energy from the atomic
IP to finite particle sizes to the bulk WF.  It is interesting to ask
whether some simple parametrizations for the image force cutoff
distance and atomic and cluster radii may be proposed so as to enable
more quantitative applications of Eq. (\ref{IP(R)}) to experimental
data. This is the subject of the sections that follow.

\section{LENGTH SCALE PARAMETERS}
\label{sec:result}

For guidance with length scaling, let us begin by plotting the
"experimental" image force cutoff distances as defined by equating
$e^2/(4d)$ to the experimentally measured bulk surface
polycrystalline work functions\cite{CRC:01}. These values are shown
in Fig. \ref{dWF}(a).
The same figure displays two quantities reflecting the size of
individual atoms: the cube root of the atomic
polarizability\cite{CRC:01} ($r_{\alpha}$) and one-half of the
nearest neighbor distance in a crystal\cite{kittel} ($r_{nn}/2$).
Although $d$ shows strong correlation with both the polarizability
radius and the nearest neighbor distance, it's consistently lower
than either one. However, by forming the inverse sum of the two, a
quantity emerges which follows the experimental values of $d$
beautifully. This parametrization has the form
\begin{equation}
1/d \approx 1/r_{\alpha} + 2/r_{nn}
\label{d}
\end{equation}
and agrees with the empirical value of the cutoff parameter to within
$\sim$ 10\% for most metallic elements, see Fig. \ref{dWF}(b). Put
another way, the WF for most metallic elements can be estimated to
better than 10\% by using this parametrization of $d$.

The coexistence of an atomic part and a bulk part in a work function
model resembles qualitatively the elegant argument in Ref.
[\onlinecite{harrison:99}] in which the WF value is calculated as
arising from the IP of a neutral atom reduced by the work done by the
image force in bringing the resulting ion back to the crystal
surface.

As a matter of fact, Fig. \ref{dWF}(a) reveals that all the size
parameters turn out to have essentially the same trend across the
periodic table, differing only in overall magnitude. We are not aware
of a quantitative theory explaining this observation for the
crystalline state, but, roughly speaking, one does expect bigger
atoms to have higher polarizabilities, larger lattice spacings in
crystal form, and lower IP together with lower WF [via Eqs.
(\ref{IPWF},\ref{IPWF2})] and therefore greater $d$ values. In fact,
remarkably linear correlations between the polarizability radius and
such properties of individual atoms as the radius of the outermost
orbital, the ionization potential, and the electronegativity have
been extensively discussed in the literature (see Refs.
\onlinecite{dmitrieva:83,nagle:90,ghanty:96} and references therein).
This strongly suggests that just a {\it single} size parameter may be
sufficient to yield an estimate for the WF.

Indeed, as shown in Fig. \ref{IPWFfig}, it turns out that the ratio
$r_{nn}/(2r_{\alpha})$ is very close to 0.65 for all metallic
elements\cite{exptavg} (the deviation from this average is within
$\sim 20\%$), and this correlation can be substituted into Eq.
(\ref{d}). In the end, therefore, one can use, e.g., only the
polarizability radius $r_{\alpha}$ in order to predict the
approximate value of the bulk surface WF.

As shown in Fig. \ref{WFPlot}, a plot of
\begin{equation}
\label{onlyalpha} W = \frac{2.54e^2}{4r_{\alpha}},
\end{equation}
i.e., Eq. (\ref{d}) with $r_{nn}/2=0.65r_{\alpha}$ [cf. Fig.
\ref{dWF}(c)] confirms that this single-parameter formula gives WF estimates
within $\sim$ 10\% of the experimental values.
As mentioned at the beginning of Sec. \ref{sec:image}, a variety of
expressions for the quantity $d$ have been proposed earlier, but the
one given here shares the virtues of needing only a single piece of
information (the atomic polarizability or the nearest neighbor
distance), being free of additional adjustable parameters, and
producing rather accurate results.

Returning to the subject of the IP/WF ratio discussed in the previous
section, Eq. (\ref{IPWF2}), we notice that even more accurate
agreement with the experimental average is obtained if we employ
\begin{equation}
\label{d/Rat}  d/R_{at} \approx r_{nn}/(2r_{\alpha}) \approx 0.65.
\end{equation}
Indeed, this is a reasonable representation of the ratio of a
characteristic crystalline length to an atomic length. This
refinement gives
\begin{equation}
I/W \approx \eta(0.65) = 1.8, \label{IPWF3}
\end{equation}
see Fig. \ref{IPWFfig}.

With estimates of both the work function WF, and the ratio IP/WF
$\approx$ 1.8, the atomic first ionization potential can be
calculated. Fig. \ref{IPplot} shows the calculated IP plotted vs. the
exact known IP \cite{CRC:01}. The accuracy is again $\pm$ 10\% for
most elements \cite{trivalent}.

Eqs. (\ref{d}) and (\ref{d/Rat}) imply the following expression for
the size parameter $R_{at}$ in the atomic limit:
\begin{equation}
\label{Rat} R_{at} \approx \frac{{r_\alpha^2 }}{{r_\alpha+r_{nn}/2}}
\approx 0.61r_{\alpha},
\end{equation}
where in the last part we again used the average experimental ratio
between $r_{nn}$ and $r_{\alpha}$.

The discussion in this section shows that a consistent and economical
description of electron removal energies both in the bulk and the
atomic limits can be achieved with the help of only a single atomic
parameter. In the next section we will consider the region in
between, i.e., the electron removal energies for clusters of a finite
number of atoms.

\section{ESTIMATE OF ALLOY WORK FUNCTIONS}
\label{sec:alloys}

The regularities discussed above have been deduced for pure metals.
It is interesting to inquire what they might imply for mixed systems
such as binary alloys.  Obviously, it is beyond the capacity of the
present approach to achieve a microscopic understanding of mixed
systems and to incorporate the possible variations in structure and
miscibility . Nevertheless, it makes sense to consider a simple
scaling argument for continuous bulk solutions.

The preceding section demonstrated that the WF can be rather
accurately parametrized as being inversely proportional either to the
polarizability radius or to the nearest-neighbor distance.  In an
alloy $A_{x}B_{1-x}$ the effective values of these length parameters
will change.  In fact, there are two natural approximations which can
be made.  One is that the lattice parameter will change linearly with
concentration.  This is known as Vegard's
law\cite{vegard:21,kiejna:96}.  Within the framework of the present
model, the alloy WF would then behave as the inverse average of the
constituents' WF's: $1/W(x)\approx x/W_{A}+(1-x)/W_{B}$.

An alternative and possibly more physical assumption is that it is
the effective polarizability that should be taken as varying
linearly with concentration.  Indeed, the polarizability of a
small volume element of the sample is approximately the weighted
average of the polarizabilities of the components.  In this case,
with $W \sim \alpha^{-1/3}$, we anticipate

\begin{equation}
\label{alloyeq}\frac{1}{[W(x)]^3} = \frac{x}{W_A^3} +
\frac{1-x}{W_B^3}
\end{equation}

Note that this relation would also be obtained if we assumed that it
was the molar volumes, rather than lattice parameters, that would
need to be averaged upon alloying (Zen's law\cite{zen:56,kiejna:96}).

Well-controlled measurements of $W(x)$, like rigorous
calculations, are not many in number: traces of surface
segregation or contamination can obliterate the basic effect. Fig.
\ref{alloyfig} illustrates the prediction of Eq. (\ref{alloyeq})
for binary alloys of silver  and gold, carefully characterized in
Ref. \onlinecite{fain:74}. The scaling curve reproduces the trend
of the data and even the correct curvature of the Ag-Au line. It
is noteworthy that predicting the sign of the deviation of $W(x)$
from linearity appears to be a difficult challenge for microscopic
theory. The one calculation which focused on such a
deviation\cite{gelatt:74} applied a self-consistent two-band
charge-transfer model to $Ag_{x}Au_{1-x}$ and was able to yield
the correct sign only when level shifts due to alloying were taken
into account. The absolute magnitude of the calculated deviation
was approximately one-third of the experimental value, and was in
fact very close to the curve in Fig. \ref{alloyfig} produced by
Eq. (\ref{alloyeq}).

For cases when the constituents' work functions differ by
relatively small amounts, such as alkali
alloys\cite{oirschot:72,malov:74}, (Eq. (\ref{alloyeq}) predicts
an essentially linear interpolation, in agreement with the data.

It is seen, therefore, that scaling estimates can be usefully
applied to the work functions of smoothly evolving alloys.  Of
course, cases when the WF is not a monotonic function of
concentration \cite{malov:74,ishii:01} call for more involved
structural and theoretical analysis.

\section{ELECTRON REMOVAL ENERGIES FOR CLUSTERS
AND HARD SPHERE PACKING} \label{sec:radius}

Having shown that the scaling law, Eq. (\ref{IP(R)}), offers good
guidance for both IP and WF, we now address the intermediate regime
of finite clusters. Therefore a workable parametrization of cluster
radius, $R_{cl}$, is needed for use in this equation.

A commonly used method to estimate the particle radius is to scale it
according to the Wigner-Seitz radius: $R_{cl}(N)=r_{s}N^{1/3}$, where
$N$ is the number of atoms. This expression, commonly augmented with
an electron spill-out term, has been used to describe many cluster
properties. However, if this formula is used in Eq. (\ref{IP(R)})
together with the expressions for $d$ obtained above, a close match
with the experimental data is not obtained. Moreover, it is desirable
to retain a form that would depend only on $N$ and on the parameters
$r_{\alpha}$ and/or $r_{nn}$, as in the atomic and bulk limits
described above.

A complementary definition of cluster radius has been proposed in
Ref. [\onlinecite{durak:chem:01,durak:**}]. The cluster was modelled
as a group of hard spheres packed so as to minimize the surface area.
The radius of a sphere circumscribing the close-packed cluster was
then substituted into Eq. (\ref{IP(R)}). By packing steel balls in a
rubber envelope, the authors found empirically that this
circumscribing sphere radius could be expressed as
\begin{equation}
\label{packing} R_{cl}(N) \approx 1.3r_{0}N^{1/3}
\end{equation}
for clusters $N\gtrsim 7$, where $r_0$ is the radius of each
individual hard sphere. For smaller clusters, the exact
circumscribing sphere radius can be derived from simple
geometry\cite{durak:chem:01}. This definition gave reasonable answers
for several metal clusters, although additional adjustable parameters
had to be employed. We found that an expression of the type
(\ref{packing}) can be derived analytically for various cluster
packing arrangement. The precise numerical coefficient varies with
the packing structure, but in all cases is not far from 1.3. The
analysis is described in the Appendix.

Based on these considerations, we formulate the following
interpolation for the parameter $R$ to be used in Eq. (\ref{IP(R)})
for the calculation of clusters' $\phi(R)$:
\begin{equation}
\label{R} R \approx {R_{cl}^2(N) \over R_{cl}(N) + r_{nn}/2}.
\end{equation}
Here $R_{cl}(N)$ is the outer, or circumscribing (as described above)
radius for a cluster with $N$ hard spheres, each of a radius equal to
the polarizability radius $r_{0}=r_{\alpha}$. As before, $r_{nn}$ is
the nearest neighbor distance of the bulk metal . For $N\gg 1$ this
expression turns into $R_{cl}$\cite{sizecomment}, while in the atomic
limit it is the same as Eq. (\ref{Rat}).

In view of the previously discussed relation between the experimental
values of $r_{nn}$ and $r_{\alpha}$, this equation can be rewritten
purely in terms of the latter quantity. In particular, for the larger
clusters for which Eq. (\ref{packing}) holds, we obtain
\begin {equation}
\label{Ralphaonly} R \approx \frac{1.3r_{\alpha}N^{1/3}}
{1+0.5N^{-1/3}}
\end{equation}
When these expressions are used together with $d$ from the previous
section in Eq. (\ref{IP(R)}) [note that for the smallest clusters it
is important to use the full image-charge formula rather than the
expansion (\ref{finitesize})], they succeed in describing the
behavior of a large fraction of elemental metallic clusters. This is
shown in Fig. \ref{clusterPlot} which compares the calculated
electron removal energies with experimental results on clusters for
which data over a sufficient size range are available. It is not
surprising that the cluster fits are not of uniform quality
\cite{badclusters}. Indeed, whereas our aim has been to reduce as
much as possible the number of input parameters, the definition of
cluster size is inherently not so "universal." The geometric
structures of clusters containing up to several hundred atoms are
neither constant nor necessarily the same as the bulk limit (see,
e.g., the reviews in Refs.
\onlinecite{haberland:book,knickelbein:99}), and the bond lengths may
generally vary both with cluster size and from the inner to the outer
layers. As a result, the formulae given in this section should be
viewed as approximations. Their utility is in the fact that they
provide a sensible interpolation between the atomic and bulk limits
discussed above, and it is satisfying that a one-parameter expression
in many cases provides not only qualitative but also quantitative
guidance.

\section{SUMMARY}
\label{sec:Summary}

The main results of this work can be summarized as follows.

(1) Starting with an expression for the electron removal energy in
terms of the image charge potential for an isolated spherical
particle and a surface cutoff parameter [Eq. \ref{IP(R)}], we
explored how this model may be consistently applied to metallic
systems ranging from bulk surfaces to finite clusters and down to
individual atoms.

(2) We showed that this approach provides a transparent physical
explanation for the empirical fact that the atomic ionization
potentials and polycrystalline work functions of the metallic
elements exhibit an almost constant ratio of $\sim 2$ over the
periodic table [Eqs. (\ref{IPWF}, \ref{IPWF3})].

(3) We found that for most elements there is a remarkably close
numerical correlation between the values of the nearest-neighbor
distance in a crystal, the cube root of the atomic polarizability,
and the image force cutoff parameter.  This correlation may be
rationalized qualitatively, but appears worthy of further study.

(4) Taking advantage of this correlation, we formulated simple
expressions for the cutoff distance [Eq. (\ref{d})] and the atomic
and cluster radii [Eqs. (\ref{Rat},\ref{R},\ref{Ralphaonly})]. They
yield good estimates for the work function, the ionization potential,
and the cluster electron removal energies by using only a {\it single
input parameter}, the atomic polarizability. No extra adjustable
parameters are required.

(5) Generalizing the scaling argument to the case of binary alloys,
we found that it can simulate the shape of the concentration
dependence of work functions of continuous bulk solutions [Eq.
(\ref{alloyeq})].

(6) We also provided an analytical derivation of the connection
between the geometric structure of a cluster of close-packed
spheres and its outer radius. It is described by an equation of
the type (\ref{packing}), but the precise numerical coefficient is
shown to depend on the packing structure.

\begin{acknowledgments}
We are grateful to Prof. Walter E. Harrison, Dr. Tomasz Durakiewicz,
and Dr. Leeor Kronik for very informative discussions, and to the
referee for useful suggestions. This work was supported by the U.S.
National Science Foundation under Grant No. PHY-0098533.
\end{acknowledgments}

\newpage

\begin{appendix}
\section*{APPENDIX: HARD SPHERE PACKING}
There exist well-defined geometries for finite systems which provide
densely packed structures corresponding to specific bulk lattices in
the large-size limit. As reviewed in detail, e.g., in Refs.
[\onlinecite{mueller:94,martin:96,smirnov:00}], these structures are
best visualized as arising from the packing of spheres. Some of the
commonly encountered ones are illustrated in Fig. \ref{polyhedrons}.
Here we derive the outer radii of these structures as a function of
the number of hard spheres, $N$. The radius is defined as
corresponding to the {\it smallest sphere which completely encloses
(circumscribes)} the cluster made up of $N$ close-packed hard
spheres, each with a radius $r_0$

\subsection{Cubeoctahedron}
The cubeoctahedron is a shape with small surface area which can be
cut out of an {\it fcc} crystal. The number of hard spheres as a
function of the number of shells $k$ is given by\cite{martin:96}
$N=(10/3) k^3 - 5 k^2 + (11/3) k - 1$. The outer radius $R$ is given by
$R = (2k -1)r_0$. For large $k$ this becomes $R \approx 2k r_0
\approx 2 (3/10)^{1/3} r_0 N^{1/3} \approx 1.339 r_0 N^{1/3}$.  The
same expression can be derived by evaluating the volume of the
circumscribing sphere relative to the sum of the volumes of the small
hard spheres\cite{wong:02}.

\subsection{Truncated hexagonal bipyramid}
The truncated bipyramid arises from an {\it hcp} lattice. The number
of hard spheres in a truncated bi-pyramid is $N = -(3/4) + (7/2) k - (21/4)
k^2 + (7/2) k^3$ for odd $k$ and $N = -1 + (7/2) k - (21/4) k^2 + (7/2) k^3$
for even $k$.  The outer radius is $R = (2k - 1)r_0$. For large $k$, $R
\approx 2(2/7)^{1/3} r_0 N^{1/3} \approx 1.317 r_0 N^{1/3}$.

\subsection{Rhombic dodecahedron}
The rhombic dodecahedron derives from {\it bcc}. The number of hard
spheres in a rhombic dodecahedron is $N = 4 k^3 - 6 k^2 + 4 k - 1 $,
and the outer radius is $R = (4/\sqrt{3}(k - 1) + 1) r_0$ For large
$k$, $R \approx 4/\sqrt{3}(1/4)^{1/3} r_0 N^{1/3} \approx 1.455 r_0
N^{1/3}$.

\subsection{Icosahedron}

The icosahedron has the highest symmetry of all discrete point
groups. Although due to it's five-fold symmetry, the icosahedron does
not form bulk crystals, it can be considered as a slightly distorted
fcc crystal. The icosahedron structure has been observed for small
clusters of inert-gas clusters, Ca, and Mg clusters (see references
in \onlinecite{martin:96}). The relationship between the number of
hard spheres and the number of shells is the same as for the
cube-octahedron\cite{martin:96}: $N=(10/3) k^3 - 5 k^2 + (11/3) k -
1$. The radius is also given by the same expression as for the
cube-octahedron $R = (2k - 1)r_0$. The resulting expression for the
cluster radius is therefore the same as for the cube-octahedron. For
large $k$ the radius is: $R \approx 2k r_0 \approx 2 (3/10)^{1/3} r_0
N^{1/3} \approx 1.339 r_0 N^{1/3}$.
\end{appendix}

\newpage

\begin{figure}
\caption{A plot of the scaling function $\eta$(d/R) governing the
variation of the electron removal energy with size, see Eq.
(\ref{phi/WF}).}
\label{fig:scalingfunction}
\end{figure}

\begin{figure}
\caption{ Diamonds: the experimental work functions\cite{CRC:01}
plotted in terms of $d=e^2/(4W)$. (a)Triangles: $r_{\alpha}$, the
cube root of the atomic polarizability; circles: $r_{nn}/2$, one-half
of the crystalline nearest neighbor distance. (b) Squares: Eq.
(\ref{d}) for the parameter $d$. (c) Circles: $r_{\alpha}/2.54$.}
\label{dWF}
\end{figure}

\begin{figure}
\caption{Stars: a plot of the ratio $r_{nn}/(2r_{\alpha})$ for the
metallic elements (the average is $\approx$ 0.65). Diamonds:
the ratio IP/WF (the average is $\approx$ 1.8).} \label{IPWFfig}
\end{figure}

\begin{figure}
\caption{A plot of metal work functions estimated by Eq.
(\ref{onlyalpha}) vs. the experimental values.  The straight lines
mark the region of $\pm$ 10\% deviation.} \label{WFPlot}
\end{figure}

\begin{figure}
\caption{A plot of atomic ionization potentials estimated by Eqs.
(\ref{onlyalpha},\ref{IPWF3}) vs. the experimental values
\cite{CRC:01}. The straight lines mark the region of $\pm$ 10\%
deviation.} \label{IPplot}
\end{figure}

\begin{figure}
\caption{Concentration dependence of the work function of Ag-Au
binary alloys.  Solid line: estimate according to Eq.
(\ref{alloyeq}), dotted line: experimental behavior
\cite{fain:74}, dashed line: linear interpolation.
}\label{alloyfig}

\begin{figure}
\caption{A comparison between the experimental electron removal
energies of small metal clusters (dots) and the model described in
the text (solid curves). The first (circled) dot in each plot is the
atomic ionization potential.  The data were adapted from Refs.
[\onlinecite{dugourd:92}] (Li),
[\onlinecite{kappes:88,persson:91,homer:91}] (Na),
[\onlinecite{saunders:85,saunders:86}] (K),
[\onlinecite{knickelbein:94}] (Y), [\onlinecite{knickelbein:90b}]
(Nb), [\onlinecite{koretsky:97}] (Mn), [\onlinecite{yang:90}] (Fe),
[\onlinecite{yang:90}] (Co), [\onlinecite{knickelbein:90c}]
(Ni), [\onlinecite{cu:92}] (Cu), [\onlinecite{jackschath:92}] (Ag),
[\onlinecite{cd:92}] (Cd), [\onlinecite{schriver:90}] (Al),
[\onlinecite{pellarin:92}] (In), [\onlinecite{yoshida:99}] (Sn), and
[\onlinecite{koretsky:98}] (Ce,Pr).} \label{clusterPlot}
\end{figure}

\begin{figure}
\label{polyhedrons} \caption{Hard-ball and polyhedral depictions of
close-packed structures: (a) Cubeoctahedron, (b) Rhombic
dodecahedron, (c) Truncated hexagonal bipyramid, (d) Icosahedron.}
\end{figure}


\begin{thebibliography}{}
\bibitem{density}Since electrons in a metal are actually delocalized, a more
rigorous justification may be to consider the WF as the work involved
in modifying the electron density distribution. Viewing the electron
cloud as spilling out to an effective boundary a distance $d$ from
the surface, the work function can be identified with the difference
in electrostatic energy between point $d$ and infinity.
\bibitem{exptavg}Within individual structural families, the average of
the ratio $r_{nn}/(2r_{\alpha})$ is 0.72 for the {\it fcc} metals,
0.66 for the {\it hcp} metals, and 0.61 for the {\it bcc} metals.
\bibitem{trivalent}It may be noted in both Figs. \ref{WFPlot} and \ref{IPplot}
that elements with three valence electrons account for a large
fraction of deviations (B-Al-Ga-In-Tl) as well as for the switched
ordering of the IPs of the sequence Sc-Y-La. The present
semi-empirical level is not designed to pinpoint the origin of such
behavior, but it is a compelling question for further research.
\bibitem{sizecomment}In the limit of large clusters which can
be thought of as bulk crystal fragments, the radii of their
circumscribing spheres should become proportional to $r_{nn}$ rather
than $r_{\alpha}$, which implies an additional numerical factor in
Eq. (\ref{R}).  However, in the small-cluster limit for which most
experimental data are available and for which we are developing a
parametrization, the use of $r_{\alpha}$ gives a better fit.  As
remarked at the end of Sec. \ref{sec:radius}, this evidently reflects the
fact that the small metal clusters, produced in beam sources, do not
necessarily possess the exact bulk geometry.
\bibitem{badclusters}Of the six elements that
deviate the most from the scaling curves, two (Li, Mn) have a ratio
$r_{nn}/(2r_{\alpha})$ which is significantly different from the
average value of 0.65 and one (Sn) does not have a closed-packed
crystal structure. Furthermore, note that even the strongest
deviations in Fig. \ref{clusterPlot} are less than 20\%.
\end{thebibliography}
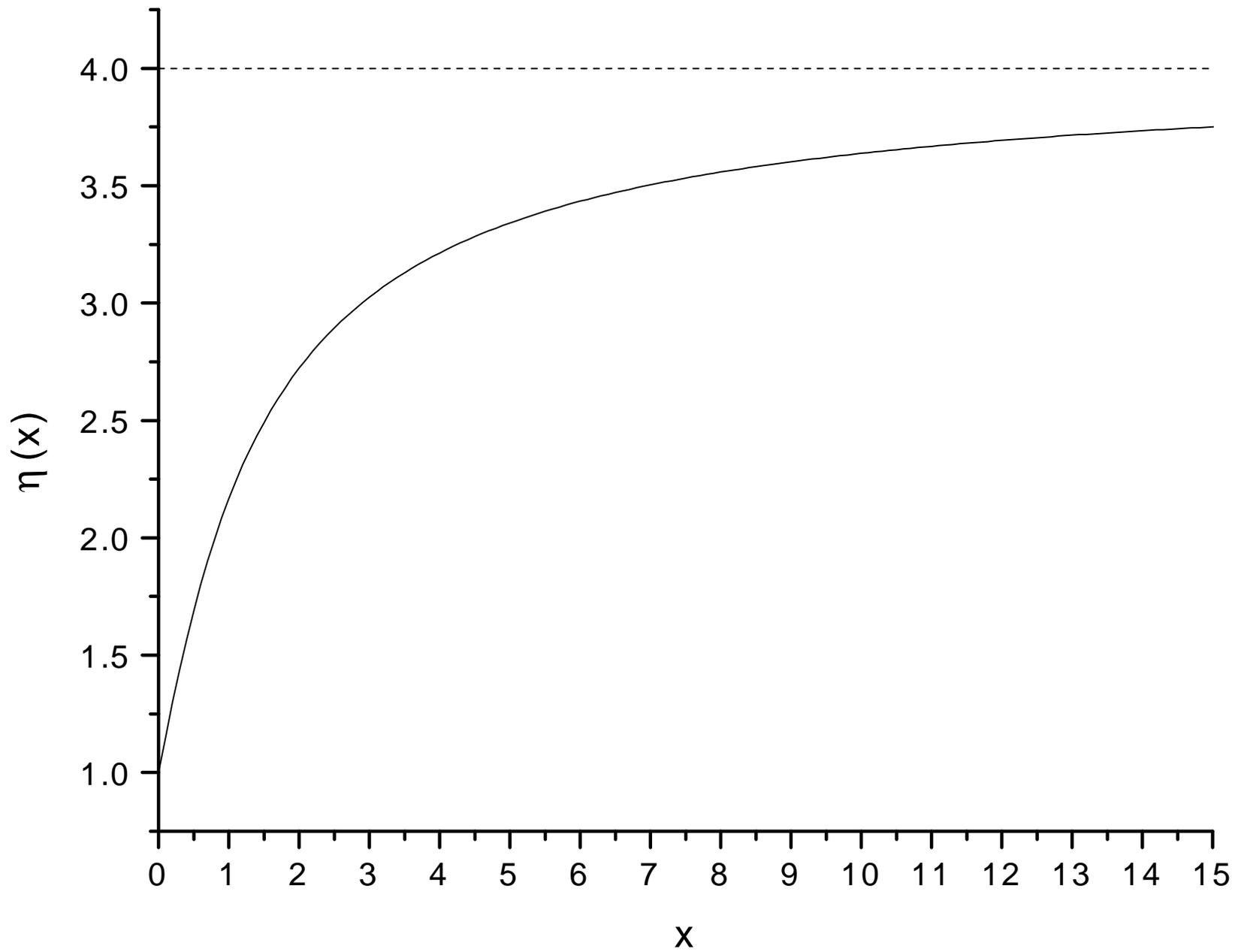
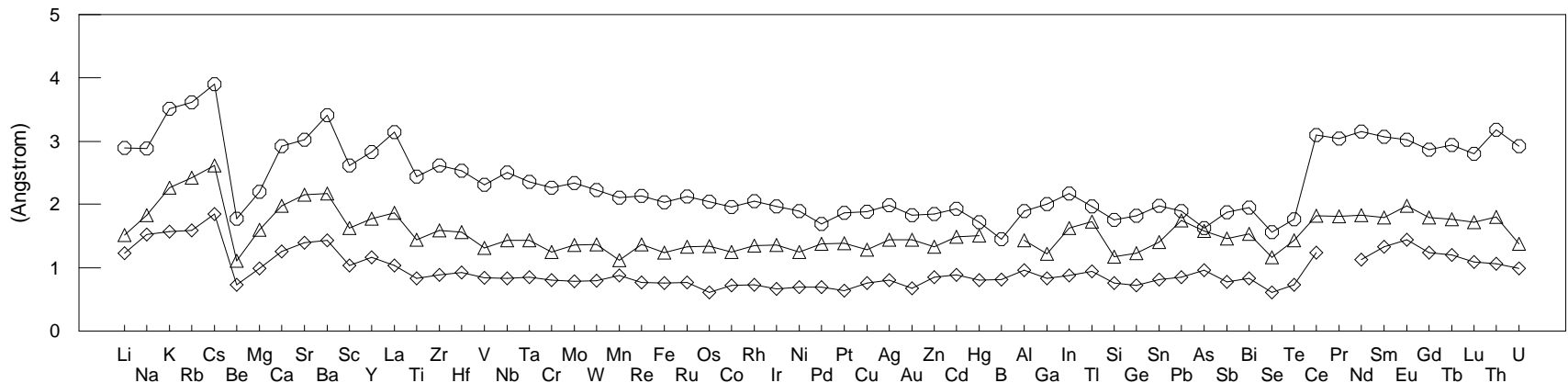
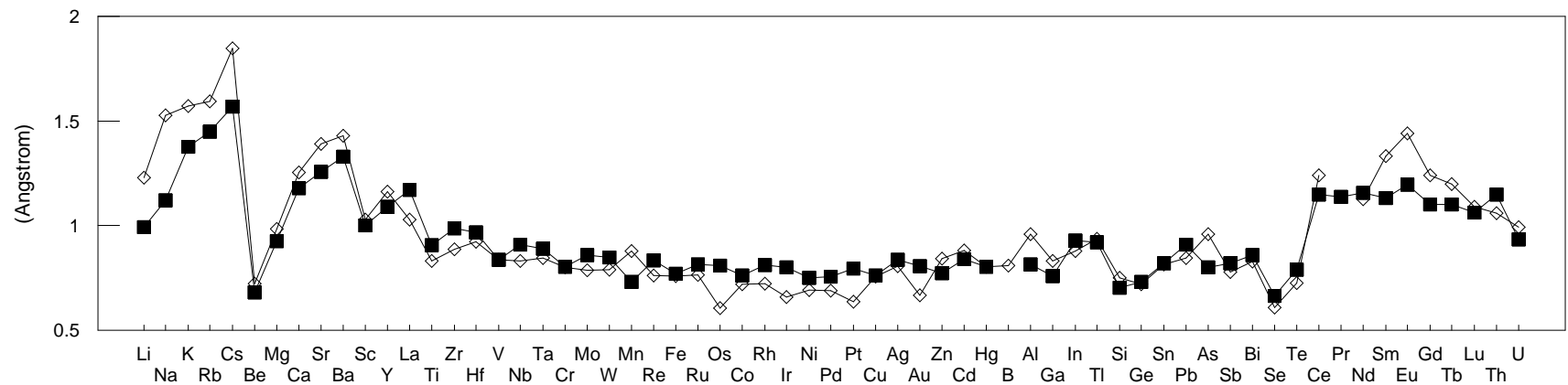
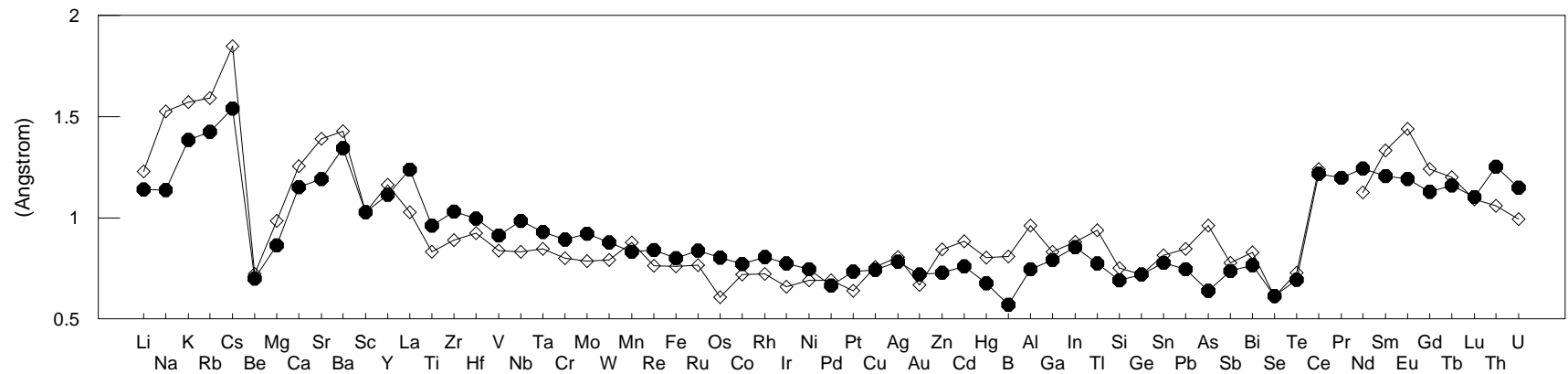
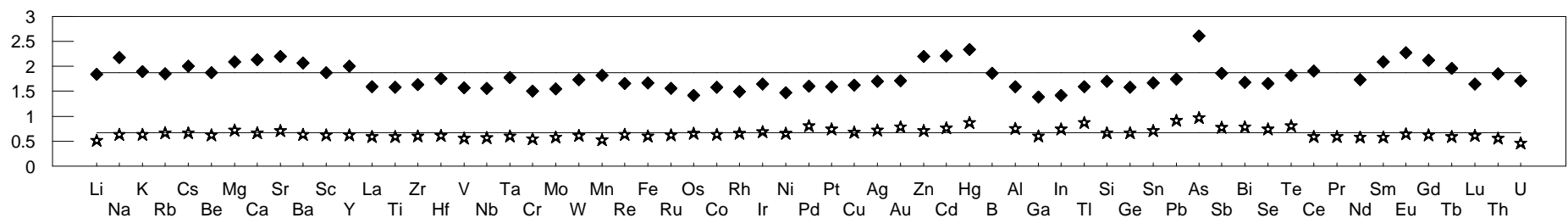
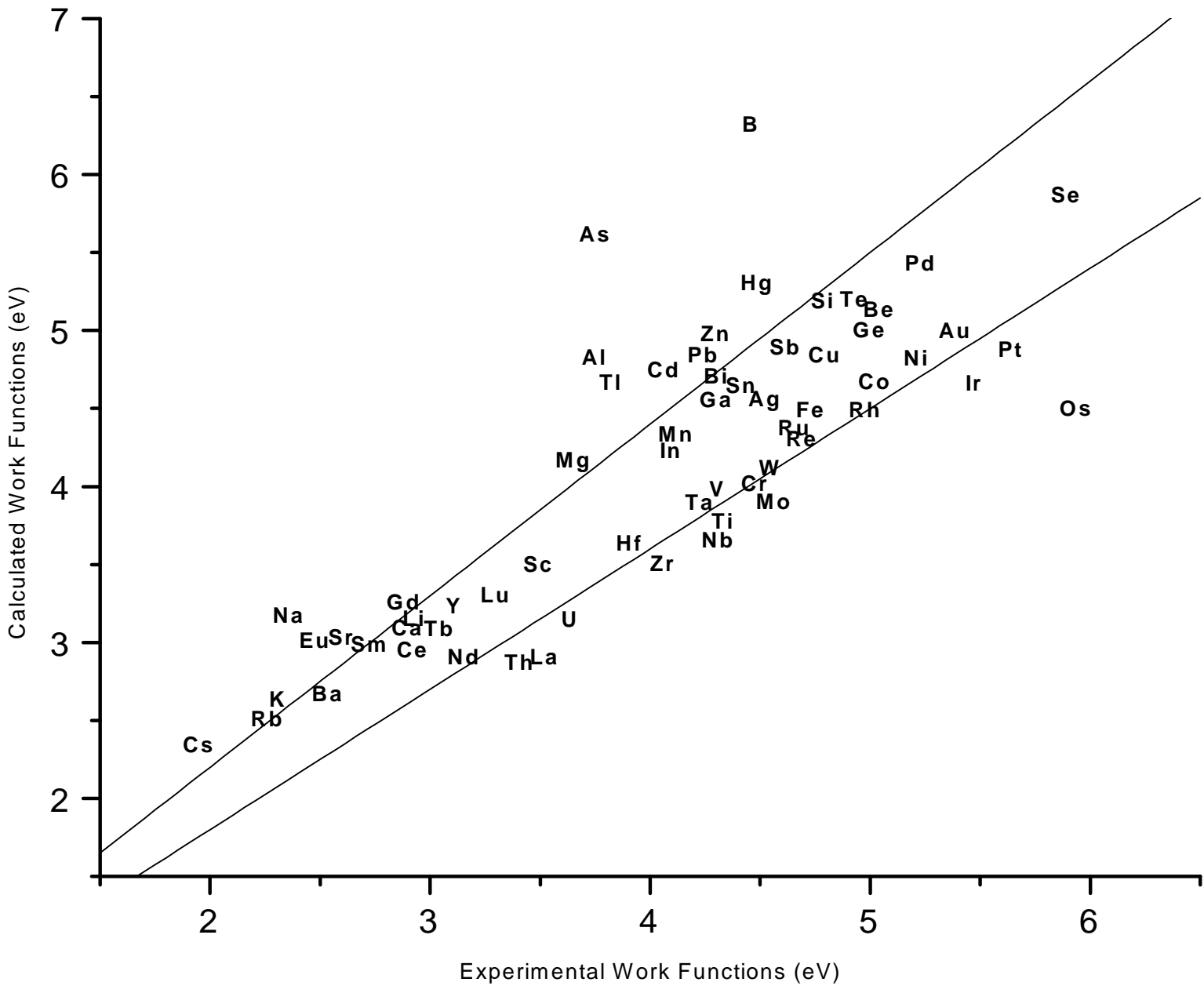
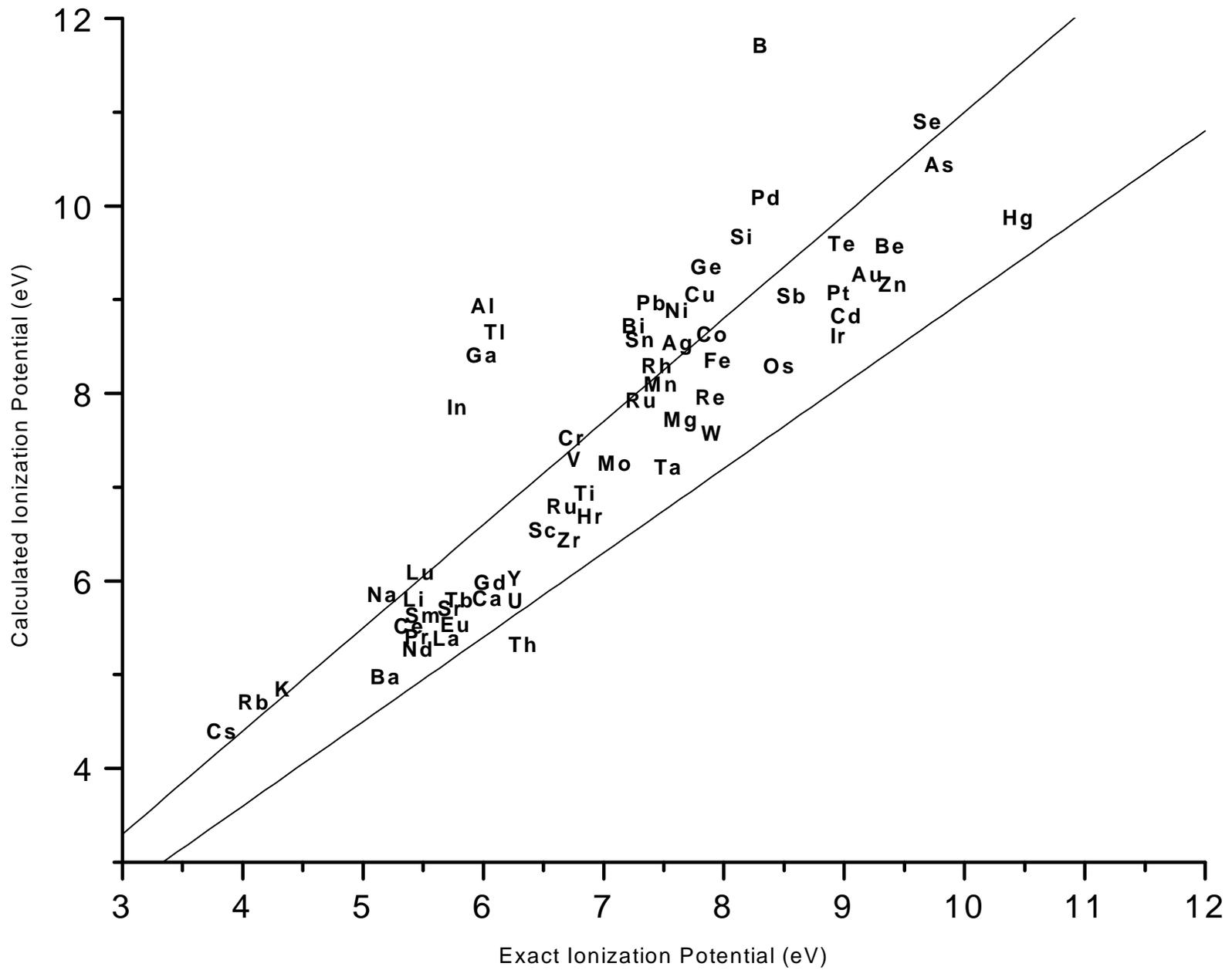
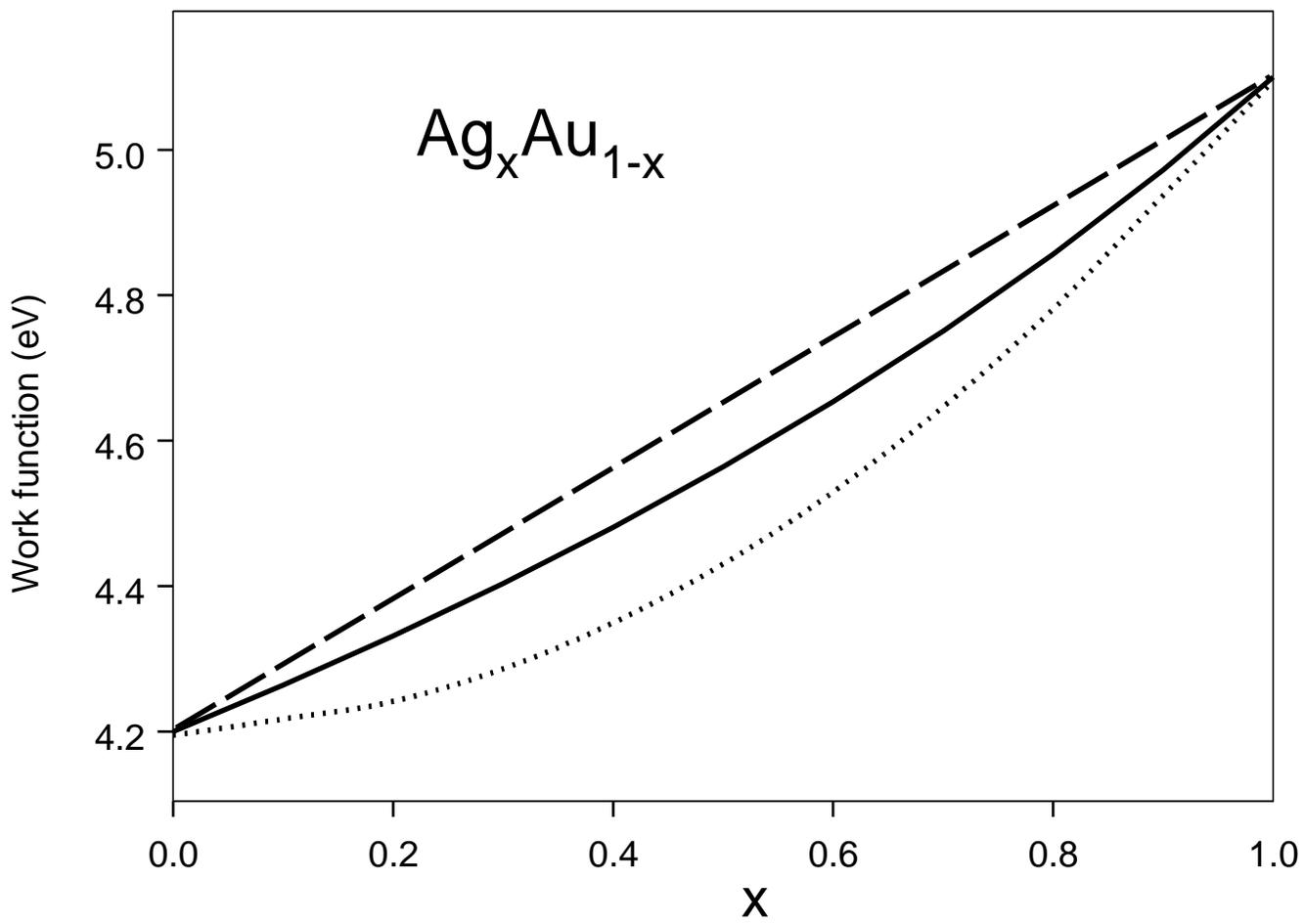
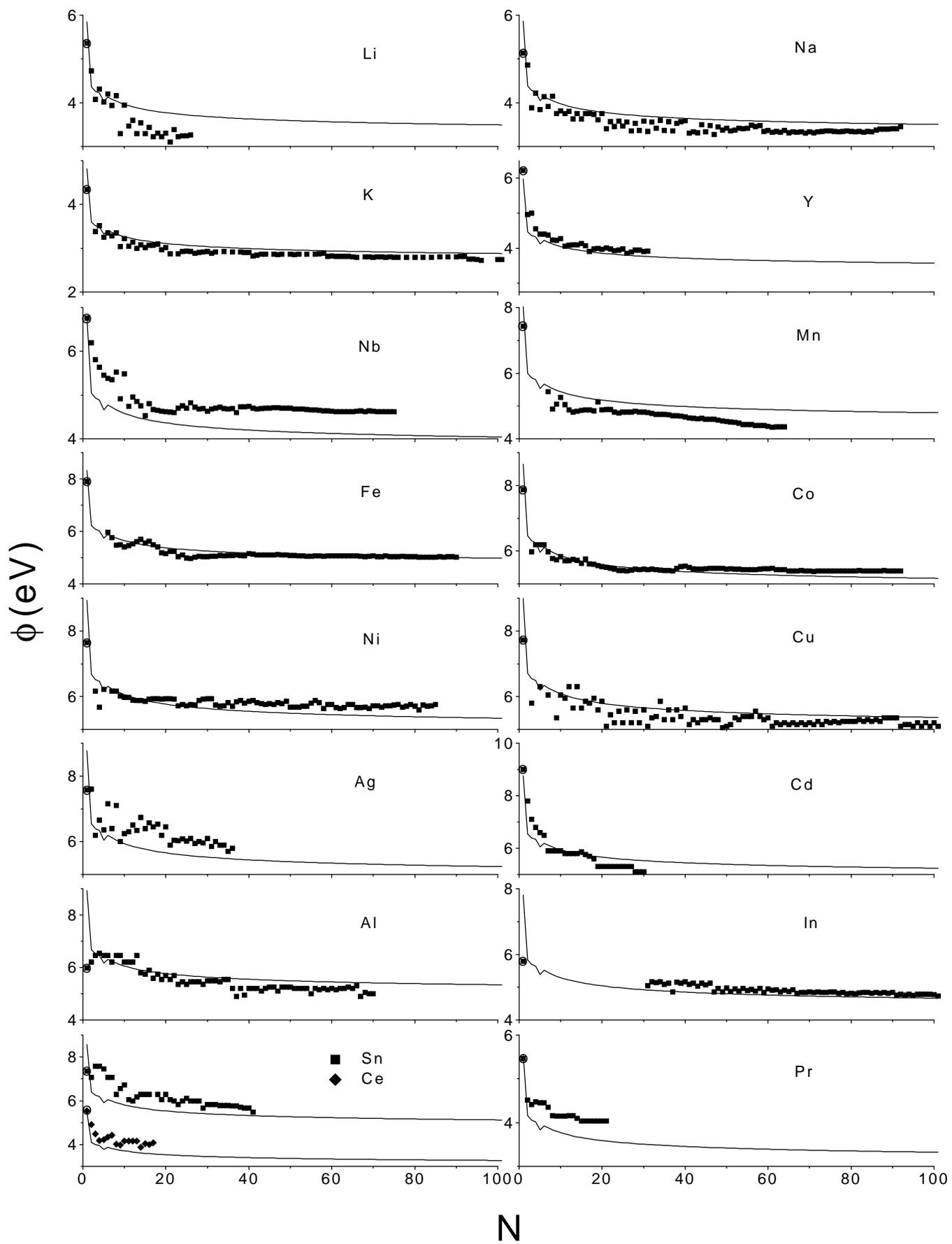
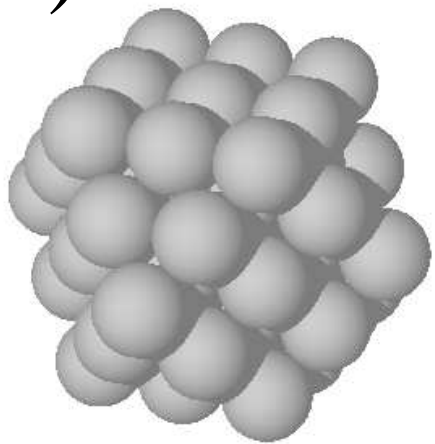
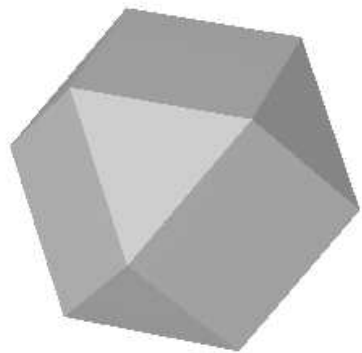
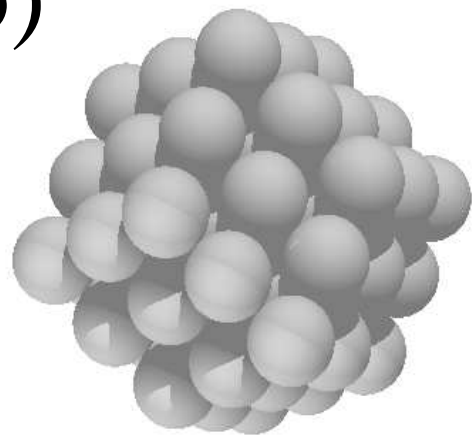
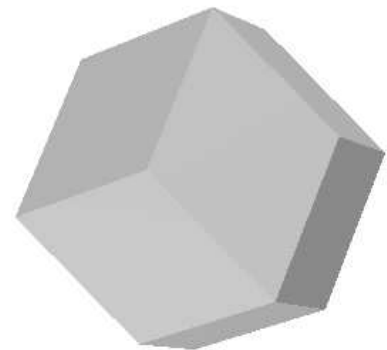
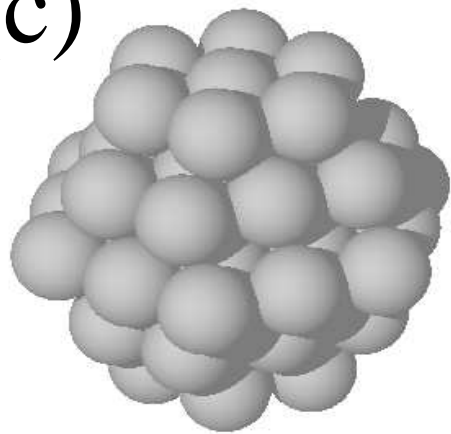
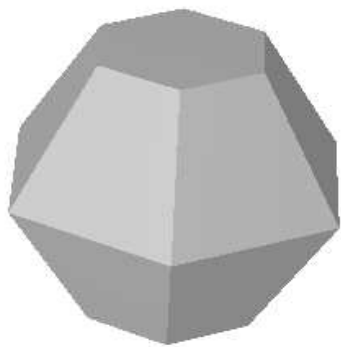
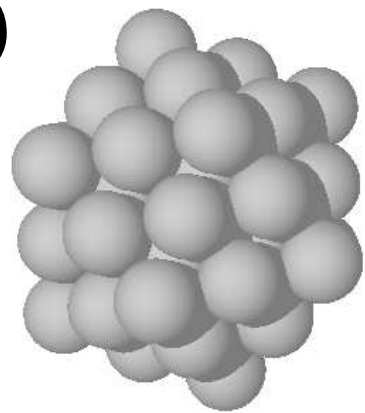
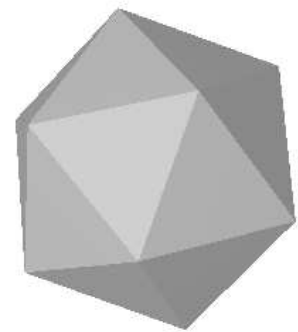
\end{document}